\documentclass[a4paper,10pt]{article}
\usepackage[utf8x]{inputenc}

\title{Scalar Field Theory on Supermanifolds}
\author{Mir Hameeda \\ Degree Collage Boys Baramulla \\
 Kashmir,  India}

\begin{document}

\maketitle

\begin{abstract}
In this paper we will analyse a scalar field theory
on  a spacetime with noncommutative and non-anticommutative coordinates. 
This will be done by using supermanifold formalism. 
We will also analyse 
its quantization in path integral formalism.  
\end{abstract}

\section{Introduction}  Path integral quantization is 
manifestly Lorentz covariant. In this method a sum over
 all field configurations
is performed.   
Path integration has proven a very useful method to quantize 
most  field Theories. 
 It is thus possible to quantize 
gauge theories and even perturbative quantum
gravity  using path integrals
    \cite{a}-\cite{f}.  However, the quantization 
of perturbative quantum gravity can only be formally done. Field theories on
spacetime with noncommutative and non-anticommutative 
coordinates have been studied using  supermanifolds formalism
\cite{sm}-\cite{smu}.  Supermanifolds have both bosonic  
and  fermionic coordinates and hence are well suited  for
 analyzing spacetime with both the noncommutative and 
non-anticommutative coordinates. There is a map between
 these noncommutative and non-anticommutative coordinates 
and ordinate coordinates.  The noncommutativity and non-anticommutativity
 induces a star product between ordinary coordinates. 
In this paper we will analyse the quantization of a scalar field theory on such supermanifolds using path integrals. 
\section{Supermanifolds}
We will now analyse a spacetime with 
 noncommutative and non-anticommutative coordinates  in 
the language of supermanifolds. 
The coordinates of the supermanifolds can be written as 
\begin{equation}
\label{supercoordinates}
z  ^\mu  =x^\mu  +  y^\mu  ,
\end{equation}
where $x^\mu  $ are the bosonic coordinates with even Grassmann parity,
\begin{equation}
[x^\mu  ,x^\nu  ]=0,
\end{equation}
and $y^\mu  $ are  the fermionic coordinates with 
odd Grassmann parity,
\begin{equation}
\{y^\mu  ,y^\nu  \}\equiv y^\mu  y^\nu  +y^\nu  y^\mu =0.
\end{equation}
So the metric can be now written as 
\begin{equation}
 ds^2 = g_{\mu \nu } dz  ^\mu   dz  ^\nu.
\end{equation}
We want to impose noncommutativity and non-anticommutativity 
in such a way that the theory reduces to the 
usual noncommutative theory when there is no non-anticommutativity 
and it also reduces the usual non-anticommutative theory when there is no 
noncommutativity. This can be done by imposing the following 
relations
\begin{eqnarray}
[{\hat z  }^\mu  ,{\hat z  }^\nu  ]&
=& 2y^\mu y^\nu  +i\theta^{\mu \nu }+O(\theta^2),
\nonumber \\ 
\{{\hat z  }^\mu  ,{\hat z  }^\nu  \}&=&2x^ \mu 
x^\nu  +2i (x^\mu  y^\nu  +x^\nu  y^\mu  ) -\tau^{\mu \nu }+O(\tau^2).
\end{eqnarray}
Now in the limit $y^\mu  \rightarrow 0$ and
$\tau^{\mu \nu  }\rightarrow 0$, we get 
\begin{eqnarray} [{\hat x}^\mu  ,{\hat
x}^\nu  ]&=&i\theta^{ \mu\nu },
\nonumber \\
 \{{\hat x}^\mu  ,{\hat
x}^\nu  \}&=&2x^\mu   x^\nu  
\end{eqnarray}
and in the limit $x^\mu \rightarrow 0$ and
 $\theta^{\mu \nu }\rightarrow 0$,
 we get 
\begin{eqnarray}
[{\hat y}^\mu ,{\hat y}^\nu  ]&=&2y^\mu  y^\nu  ,
\nonumber \\ 
\{{\hat y}^\mu  ,{\hat y}^\nu  \}&=&\tau^{\mu \nu }.
\end{eqnarray}

A matrix valued superfields $\hat \phi (\hat{z}) $ on this
 deformed superspace.
We use Weyl
ordering and  express the Fourier transformation of this superfield as, 
\begin{equation}
\hat \phi (\hat{z}) =
\int d^4 k e^{-i k \hat{z} } \;
\phi (k).
\end{equation}
Now  we  have a one to one map between a function of
$ \hat{z}$ to a function of ordinary
 superspace coordinates $z$ via
\begin{equation}
\phi (z)  =
\int d^4 k  e^{-i k z } \;
\phi (k).
\end{equation}
 We can express the product of two fields  
${\hat \phi}(\hat{z}) { \hat{ \phi} } (\hat{z})$
on this deformed superspace as
\begin{eqnarray}
{\hat \phi}(\hat{z}) { \hat{ \phi}} 
 (\hat{z}) &=&
\int d^4 k_1 d^4 k_2
\exp -i( ( k_1 +k_2) z ) \nonumber \\
&& \times   \exp  (
\omega^{\mu\nu} k^2_\mu k _\nu^1 )
{ \phi}  (k_1) { \phi}   (k_2),
\end{eqnarray}
where $\omega^{\mu\nu  }$ is a nonsymmetric tensor
\begin{equation} 
\omega^{\mu\nu  }=\tau^{\mu\nu} +\theta^{\mu\nu}.
\end{equation}
So we can now define the star product  between ordinary 
functions  as follows:
\begin{eqnarray}
{\phi}(z) * {  \phi}  (z) =
 \exp  (
\omega^{\mu\nu} k^2_\mu k _\nu^1 ) 
 {\phi}(z_1) {  \phi_{ }}  (z_2)
\left. \right|_{z_1=z_2=z}.&&
\label{star2}
\end{eqnarray}

\section{Scalar Field Theory}
In any field theory
the classical action $S$ is the spacetime integral 
of the Lagrangian density $\mathcal{L}$
 \begin{equation}
  S = \int d^4 x \mathcal{L}.
 \end{equation}
The exact form of $\mathcal{L}$ depends on the theory under consideration.
In scalar field theory the Lagrangian $\mathcal{L}$ is taken
 to be a function of field $\phi$ and its derivative $\partial _c \phi$. 
Usually no higher derivative terms are taken as they  lead to a 
non-unitary behavior of the field theory. 
Now we take the following action  for   free scalar field theory
\begin{equation}
 \mathcal{L} = \frac{1}{2}(\partial _a\phi *
 \partial ^a\phi + m^2 \phi * \phi),
\end{equation}
where $m$ is the mass of the field theory.

Now if we assume that the field vanishes on boundaries,
 then we can show that the action can be written as follows:
 \begin{equation}
  S = \int d^4x  
\frac{- 1}{2}\phi( \partial ^2 - m^2 ) *\phi.
 \end{equation}
This is obtained by integrating by parts and neglecting 
the surface terms. It is important that the fields vanish 
on the boundaries otherwise we will get a term proportional 
to the boundary value of the fields. If for example 
$ \tilde\phi $ is the boundary value of $\phi $
and the boundary is along $x^3$ direction, then 
\begin{equation}
\int d^4x \partial _a\phi *\partial ^a\phi =- \int d^4x  \phi
 \partial ^2  *\phi +\int d^4x  \partial_a (\phi  \partial ^a *\phi ),
\end{equation}
but the term $\partial_a (\phi \partial ^a * \phi )$
 now equals to a boundary term given by 
 $ \tilde \phi \partial ^3  * \tilde\phi$.
However,  in all our discussions we will assume
 that all the fields vanish on the boundary and hence drop any 
terms which are total derivatives as they can be converted 
into a boundary term. We can quantize this theory by first writing 
down the partition function which is give by 
 \begin{equation}
  Z = N\int D\phi \exp i [S ].
 \end{equation}
Now if $J (x)$ is the source of the field 
$\phi (x)$, then we define $J*\phi$ as 
 \begin{equation}
  J*\phi = \int d^4x J(x) *\phi(x).
 \end{equation}
We call $Z[J]$  the vacuum-to-vacuum 
 transition amplitude and it is given by
 \begin{equation}
  Z[J] = N\int D\phi \exp i [S + J*\phi ].
 \end{equation}
It is the generator of two point functions. 
 Here $N$ is a normalization constant, 
such that $Z[J]$ is normalized as follows:
 \begin{equation}
  Z[0] =1.
 \end{equation}
It can be shown that  $Z[J]$ is given by 
\begin{equation}
 Z[J] = \exp \frac{-i}{2} \int d^4x d^4x' J(x) G(x,x')J(x'),
\end{equation}
where $G(x,x')$ is Green's function, which can be viewed as the inverse of $E(x, x')$,
\begin{equation}
( \partial ^2 - m^2 ) G(x',x'') =  \delta (x',x'').
\end{equation}
Then the two-point function  is given by
\begin{equation}
- \frac{\delta}{\delta J(x)}\frac{\delta}{\delta J(x')}
 Z[J]\left. \right|_{J= 0} = i G(x, x').
\end{equation}
Any higher powers of the field variable are viewed as  
interactions. Interaction modifies $Z[J]$, but apart 
from that the procedure of calculating the two-point function  
remains identical to the free field case. This completes our 
discussion of calculating the two-point function
  for scalar field theory.

\section{Conclusion}
In this paper we have analysed the quantization of a scalar field theory on 
noncommutative and non-anticommutative spacetime. 
We first analysed the theory on supermanifolds.
Then we quantized it  by using path integration. 
It will  be interesting to analyse gauge theories 
in this spacetime. 
It may be noted that gauge symmetry  also occurs in the ABJM and 
BLG theories. The ghost and gauge fixing terms of these theories 
has been studied \cite{7a}-\cite{8a}. It will 
be interesting to study the ABJM and the BLG theories 
 on supermanifolds. 
It may also be noted that another way to deal this problem is the
Wheeler-DeWitt approach \cite{wd}-\cite{wd1}.

\end{document}